\documentstyle[12pt]{article}
\newcommand{\eut}{\;
\begin{picture}(3,3)(-2,-2)
\put(-2,-15){$\tilde{}$}
\put(-5,-2){$\eta$}
\end{picture}}

\newcommand{\ju}{
\begin{picture}(9,5)(0,0)
\put(3,12){\tiny 1}
\put(0,0){$\tilde{\Psi}$}
\end{picture}}

\newcommand{\jd}{
\begin{picture}(9,5)(0,0)
\put(3,12){\tiny 2}
\put(0,0){$\tilde{\Psi}$}
\end{picture}}

\newcommand{\fu}{
\begin{picture}(9,5)(0,0)
\put(2.5,10){\tiny 1}
\put(0,0){F}
\end{picture}}

\newcommand{\fd}{
\begin{picture}(9,5)(0,0)
\put(2.5,10){\tiny 2}
\put(0,0){F}
\end{picture}}

\newcommand{\bu}{
\begin{picture}(9,5)(0,0)
\put(2.5,12){\tiny 1}
\put(0,0){$\tilde{{\rm B}}$}
\end{picture}}

\newcommand{\bd}{
\begin{picture}(9,5)(0,0)
\put(2.5,12){\tiny 2}
\put(0,0){$\tilde{{\rm B}}$}
\end{picture}}

\newcommand{\au}{
\begin{picture}(9,5)(0,0)
\put(3,10){\tiny 1}
\put(0,0){A}
\end{picture}}

\newcommand{\ad}{
\begin{picture}(9,5)(0,0)
\put(2.5,10){\tiny 2}
\put(0,0){A}
\end{picture}}

\begin{document}

\hspace{11.6cm} CGPG-94/11-3

\begin{center}

\baselineskip=24pt plus 0.2pt minus 0.2pt
\lineskip=22pt plus 0.2pt minus 0.2pt

 \Large

Solving the Constraints of General Relativity\\

\vspace*{0.35in}

\large

J.\ Fernando\ Barbero\ G. $^{\ast, \dag}$
\vspace*{0.25in}

\normalsize

$^{\ast}$Center for Gravitational Physics and Geometry,\\
Department of Physics,\\
Pennsylvania State University,\\
University Park, PA 16802\\
U.S.A.\\

$^{\dag}$Instituto de Matem\'aticas y F\'{\i}sica Fundamental, \\
C.S.I.C.\\
Serrano 119--123, 28006 Madrid, Spain
\\

\vspace{.5in}
November 7, 1994\\
\vspace{.5in}

ABSTRACT

\end{center}
\vspace{.5in}

I show in this letter that it is possible to solve some of the constraints
of the $SO(3)$-ADM formalism for general relativity by using an approach
similar to the one introduced by Capovilla, Dell and Jacobson to solve the
vector and scalar constraints in the Ashtekar variables framework. I
discuss the advantages of this approach and compare it with similar
proposals for different Hamiltonian formulations of general relativity.

\pagebreak

\baselineskip=24pt plus 0.2pt minus 0.2pt
\lineskip=22pt plus 0.2pt minus 0.2pt

\setcounter{page}{1}

The main purpose of this letter is to show that it is possible to solve
some of the constraints of general relativity in the $SO(3)$-ADM formalism
\cite{ABJ} by using techniques similar to those introduced by Capovilla,
Dell and Jacobson \cite{CDJ} to solve the vector and scalar constraints in
terms of Ashtekar variables. More specifically, I will find that it is
possible to solve both the ``Gauss law" (the generator of the internal
$SO(3)$ rotations) and the scalar constraint by introducing suitable
$3\times 3$ $SO(3)$ matrices and imposing simple conditions on them. This
result shows that the first step in Ashtekar's procedure to derive the new
variables formalism --the introduction of an internal $SO(3)$ symmetry--
has interesting ramifications even outside the Ashtekar variables
framework. Also, it suggests some interesting relationships between three
dimensional diffeomorphisms and the $SO(3)$ transformations. Finally it
may be useful for numerical relativists because it reduces the number of
constraint equations that must be solved, it does not require the
introduction of reality conditions for Lorentzian signature space times
and the evolution equations can be written in terms of the new fields that
will be introduced later.

The conventions and notation used throughout the letter are the following.
Tangent space indices and $SO(3)$ indices are represented by lowercase
Latin letters from the beginning and the middle of the alphabet
respectively. The 3-dimensional Levi-Civita tensor density and its inverse
are denoted\footnote{I represent the density weights by the usual
convention of using tildes above and below the fields.} by
$\tilde{\eta}^{abc}$ and $\eut_{abc}$ and the internal $SO(3)$ Levi-Civita
tensor by $\epsilon_{ijk}$. The variables in the $SO(3)$-ADM phase space
are a densitized triad $\tilde{E}_{i}^{a}$ (with determinant denoted by
$\tilde{\!\tilde{E}}$) and its canonically conjugate object $K_{a}^{i}$
(closely related to the extrinsic curvature). The (densitized) three
dimensional metric built from the triad is denoted
$\tilde{\;\;\tilde{q}^{ab}}\equiv \tilde{E}^{a}_{i}\tilde{E}^{bi}$ and its
determinant $\tilde{\!\tilde{q}}$ so that
$q^{ab}=\frac{\tilde{\;\;\;\tilde{q}^{ab}}}{ \;\tilde{\!\tilde{q}}}$.  I
will use also the $SO(3)$ connection $\Gamma_{a}^{i}$ compatible with the
triad.  The variables in the Ashtekar phase space are $\tilde{E}_{i}^{a}$,
again, and the $SO(3)$ connection $A_{a}^{i}$. The curvatures of
$A_{a}^{i}$ and $\Gamma_{a}^{i}$ are respectively given by
$F_{ab}^{i}\equiv 2\partial_{[a} A_{b]}^{i}+\epsilon^{i}_{\;\;jk}A_{a}^{j}
A_{b}^{k}$ and $R_{ab}^{i}\equiv 2\partial_{[a}
\Gamma_{b]}^{i}+\epsilon^{i}_{\;\;jk}\Gamma_{a}^{j} \Gamma_{b}^{k}$.
Finally, the action of the covariant derivatives defined by these
connections on internal indices are\footnote {They may be extended to act
on tangent indices by introducing a space-time torsion-free connection;
for example the Christoffel symbols $\Gamma_{ab}^{c}$ built from $q^{ab}$.
All the results presented in the paper will be independent of such
extension}
$\nabla_{a}\lambda_{i}=\partial_{a}\lambda_{i}+\epsilon_{ijk}A_{a}^{j}
\lambda^{k}$ and ${\cal
D}_{a}\lambda_{i}=\partial_{a}\lambda_{i}+\epsilon_{ijk}\Gamma_{a}^{j}
\lambda^{k}$.  The compatibility of $\Gamma_{a}^{i}$ and $\tilde
{E}_{i}^{a}$ means ${\cal
D}_{a}\tilde{E}_{i}^{b}\equiv\partial_{a}\tilde{E}^{b}_{i}+
\epsilon_{i}^{\;\;jk}\Gamma_{aj}
\tilde{E}^{b}_{k}+\Gamma_{ac}^{b}\tilde{E}^{c}_{i}-\Gamma_{ac}^{c}
\tilde{E}^{b}_{i}=0$.

The $SO(3)$-ADM constraints are \cite{ABJ}
\begin{eqnarray}
& & \epsilon_{ijk}K_{a}^{j}\tilde{E}^{ak}=0\label{1}\\
& & {\cal D}_{a}\left[\tilde{E}^{a}_{k}K_{b}^{k}-\delta_{b}^{a}
\tilde{E}^{c}_{k}
K_{c}^{k}\right]=0\label{2}\\
& & -\zeta\sqrt{\tilde{\!\tilde{q}}}R+\frac{2}{\sqrt{\tilde{\!\tilde{q}}}}
\tilde{E}^{[c}_{k}\tilde{E}^{d]}_{l}K_{c}^{k}K_{d}^{l}=0\label{3}
\end{eqnarray}
where $R$ is the scalar curvature of the three-metric $q_{ab}$ (the inverse of
$q^{ab}$).
and the variables $K_{ai}(x)$ and $\tilde{E}^{b}_{j}(y)$ are canonical;
i.e. they satisfy
\begin{eqnarray}
& & \left\{K_{a}^{i}(x), K_{b}^{j}(y)\right\}=0\nonumber\\
& & \left\{\tilde{E}^{a}_{i}(x),
 K_{b}^{j}(y)\right\}=\delta_{j}^{i}\delta_{a}^{b}\delta^{3}(x,y)
\label{111}\\
& &  \left\{\tilde{E}^{a}_{i}(x), \tilde{E}^{b}_{j}(y)\right\}=0\nonumber
\end{eqnarray}
By choosing  $\zeta=+1$ or $\zeta=-1$ we can describe
both Lorentzian and Euclidean signature space-times. The constraints
(\ref{1}-\ref{3}) generate internal $SO(3)$ rotations, diffeomorphisms
and time evolution respectively.

For non-degenerate metrics we can multiply (\ref{3}) by
$\sqrt{\tilde{\!\tilde {q}}}$  to get
\begin{equation}
 -\zeta\;\tilde{\!\tilde{q}}R+2
\tilde{E}^{[c}_{k}\tilde{E}^{d]}_{l}K_{c}^{k}K_{d}^{l}=0
\label{4}
\end{equation}
Using now
\begin{equation}
\tilde{\!\tilde{q}}R=-\epsilon^{ijk}\tilde{E}^{a}_{i}\tilde{E}^{b}_{j}R_{abk}
\label{5}
\end{equation}
equation (\ref{4}) can be rewritten in the form
\begin{equation}
\epsilon^{ijk}\tilde{E}^{a}_{i}\tilde{E}^{b}_{j}\left(\zeta
R_{abk}+\epsilon_{k}^{\;\;lm}K_{al}K_{bm}\right)=0
\label{6}
\end{equation}
We follow now a procedure very closely related to the one used by
Capovilla, Dell and Jacobson to solve the scalar and vector constraints in
the Ashtekar formalism. First we define the $3\times 3$ matrix $\Psi_{ij}$
(for non-degenerate triads) as
\begin{equation}
\tilde{E}^{ai}\Psi_{ij}\equiv\tilde{\eta}^{abc}\left(\zeta
R_{bcj}+\epsilon_{jkl}K_{b}^{k}K_{c}^{l}\right)
\label{7}
\end{equation}
Introducing (\ref{7}) in (\ref{6}) we get immediately the condition
$tr \Psi=0$. We consider now the Gauss law Eq.(\ref{1}). In order to
solve it we need to write
$K_{ai}$ as a function of $\tilde{E}^{a}_{i}$ and $\Psi_{ij}$. Defining
\begin{equation}
\tilde{S}^{a}_{i}\equiv\tilde{E}^{aj}\Psi_{ji}-\zeta \tilde{\eta}^{abc}R_{bci}
\label{8}
\end{equation}
we have
\begin{equation}
K_{a}^{i}=\frac{1}{2\sqrt{2\;\tilde{\!\tilde{S}}}}\;\eut_{abc}\epsilon^{ijk}
\tilde{S}^{b}_{j}\tilde{S}^{c}_{k}
\label{9}
\end{equation}
where $\tilde{\!\tilde{S}}$ is the determinant of $\tilde{S}^{a}_{i}$. This
last expression is only valid when $\tilde{\!\tilde{S}}\neq 0$ (or
equivalently $\det K\neq 0$). Introducing now (\ref{9})
in the Gauss law (\ref{1}) we get
\begin{equation}
\eut_{abc}\tilde{S}^{b}_{j}\tilde{S}^{c}_{k}
\tilde{E}^{aj}=0
\label{10}
\end{equation}
Multiplying this last expression by
$\eut_{def}\epsilon^{kpq}\tilde{S}^{d}_{p}\tilde{S}^{e}_{q}$ and taking into
account, again, that we require $\tilde{\!\tilde{S}}\neq 0$ we find the
equivalent condition
\begin{equation}
\tilde{E}^{[a}_{i}\tilde{S}^{b]i}=0
\label{11}
\end{equation}
By using now the Bianchi identity $\tilde{E}^{a}_{i}R_{ab}^{i}=0$ we
finally obtain $\Psi_{[ij]}=0$. The results derived above mean that by
taking a symmetric and traceless $\Psi_{ij}$ it is possible to solve both
the Gauss law and the scalar constraint. The only equation left to solve
is the vector constraint. This is a nice result; we can say that, although
introducing an internal $SO(3)$ symmetry seems to complicate the theory
unnecessarily, using the previous reasoning not only we can solve the
additional constraint that generates the new internal symmetry, but also
the scalar constraint.

We can write the vector constraint in terms of $\tilde{E}^{a}_{i}$ and
$\Psi_{ij}$ by using (\ref{9}) and multiplying by $\tilde{\!\tilde{S}}^{3/2}$.
Proceeding in this way we find that it is equivalent to
\begin{equation}
\tilde{E}^{a}_{i}\tilde{S}^{d}_{k}\tilde{S}^{c_{1}}_{k_{1}}
\tilde{S}^{c_{2}}_{k_{2}}
\tilde{S}^{c_{3}}_{k_{3}}\left(4\eut_{c_{1}c_{2}c_{3}}\eut_{cd[a}
\epsilon^{k_{1}k_{2}
k_{3}}\epsilon^{ijk}-3\eut_{c_{1}c_{2}c}\eut_{c_{3}d[a}\epsilon^{k_{1}k_{2}
j}\epsilon^{ik_{3}k}\right){\cal D}_{b]}\tilde{S}^{c}_{j}=0
\label{12}
\end{equation}
where we must now write $\tilde{S}^{a}_{i}$ in terms of
$\tilde{E}^{a}_{i}$ and $\Psi_{ij}$. As we can see this is a complicated
expression (although some simplification may be achieved by imposing that
$\Psi_{ij}$ must be symmetric and traceless). It is a third order partial
differential equation in the triad fields (due to the derivatives of the
curvature) and first order in $\Psi_{ij}$. In spite of not being a simple
expression, Eq. (\ref{12}) has some features that make it interesting .
First of all it is written explicitly in terms of $\Psi_{ij}$ and
$\tilde{E}^{a}_{i}$ only; --in fact it is a polynomial equation in this
variables-- In some other cases (that I will discuss later) where some of
the constraints can be solved by using a procedure very similar to the one
described above, the equation that is left cannot be simply written in
terms of $\Psi_{ij}$ and $\tilde{E}^{a}_{i}$ (although this can be
achieved, in principle, by solving an additional differential equation).
Second, Eq. (\ref{12}) can be used for both Lorentzian and Euclidean
signatures just by selecting $\zeta=-1$ or $\zeta=+1$. This is an
advantage over some related results because there is no need to implement
the reality conditions\footnote{This fact is one of the reasons that seem
to have prevented the use of the neat idea of Capovilla, Dell and Jacobson
in numerical relativity.} . Actually, it is probably fair to say that this
is the way to explicitly incorporate the reality conditions in this type
of solution to the constraints of general relativity.

In the following I will compare the result derived above with some
closely related approaches to solve the constraints of general relativity
in different Hamiltonian formulations. I consider first the familiar
Capovilla-Dell-Jacobson approach \cite{CDJ}. Starting from the constraints
of general relativity in the Ashtekar formulation \begin{eqnarray} & &
\nabla_{a}\tilde{E}^{a}_{i}=0\label{13}\\ & &
F_{ab}^{i}\tilde{E}^{b}_{i}=0\label{14}\\ & &
\epsilon^{ijk}\tilde{E}^{a}_{i}\tilde{E}^{b}_{j}F_{abk}=0\label{15}
\end{eqnarray} they make the ansatz \begin{equation}
\Psi_{ij}\tilde{E}^{a}_{j}=\tilde{\eta}^{abc}F_{bci} \label{16}
\end{equation} Introducing this in (\ref{13}) and (\ref{14}) they
immediately get, for non-degenerate triads, that $\Psi_{ij}$ must be
symmetric and traceless. The Gauss law gives then the equation
\begin{equation}
(\nabla_{a}\Psi^{-1}_{ij})\tilde{\eta}^{abc}F_{bcj}=0
\label{17}
\end{equation}
where use has been made of the Bianchi identity
$\nabla_{a}(\tilde{\eta}^{abc}F_{bcj})=0$. If we compare this result with
the one presented in this letter we notice several interesting things.
First of all, we can solve the scalar constraint in both cases, but in one
of them we solve also the Gauss law whereas in the other we solve the
vector constraint. This suggests some hidden relationship between three
dimensional diffeomorphisms and internal $SO(3)$ gauge transformations.
Second, even though (\ref{17}) is simple in the Euclidean case, the more
relevant Lorentzian signature case is more difficult to deal with because
of the reality conditions (whose implementation is not straightforward).
In this respect our formulation is interesting because it is valid for
both Euclidean and Lorentzian space-times and no reality conditions need
to be taken into account. The equations that must be solved in both cases
are very similar because they differ only in the value of the parameter
$\zeta$.

Let us consider now the real formulation in terms of Ashtekar variables
for Lorentzian space-times discussed in \cite{fer1}. The phase space of
that formulation is the usual Ashtekar phase space but the fields are now
real. The Gauss law and the vector constraints are still given by
(\ref{13}) and (\ref{14}), whereas the scalar constraint can be written as
\begin{equation}
\epsilon^{ijk}\tilde{E}^{a}_{i}\tilde{E}^{b}_{j}(F_{abk}-2R_{abk})=0
\label{18}
\end{equation}
Making the ansatz
\begin{equation}
\Psi_{ij}\tilde{E}^{c}_{j}=\tilde{\eta}^{abc}(F_{abi}-2R_{abi})
\label{19}
\end{equation}
the vector and scalar constraints are solved if $\Psi_{ij}$ is chosen to
be symmetric and traceless as in the usual Capovilla-Dell-Jacobson case.
The Gauss law gives the equation
\begin{equation}
\nabla_{c}\left[\Psi^{-1}_{ij}\tilde{\eta}^{abc}(F_{abj}-2R_{abj})\right]=0
\label{20}
\end{equation}
Since we want to have an equation written only in terms of $\Psi_{ij}$ and
$A_{aj}$ we must remove from (\ref{20}) the terms involving the triads
$\tilde{E}^{a}_{i}$ --i.e. the term $\tilde{\eta}^{abc}R_{abk}$-- by using
(\ref{19}). In practice this amounts to solving a system of coupled
partial differential equations. Comparing this result with the one
presented in this letter we see, again, that it is the vector constraint
and not the Gauss law that is solved by choosing a symmetric $\Psi_{ij}$
and also that we must solve a system of partial differential equations
instead of the single equation (\ref{12}).

Finally I consider the two connection formulation of general relativity
discussed in \cite{fer}. The phase space in that formulation is spanned by
two $SO(3)$ connections $\au^{i}_{a}$ and $\ad^{i}_{a}$ with curvatures
given by $\bu^{a}_{i}\equiv \eta^{abc}\fu_{bci}$ and $\bd^{a}_{i}\equiv
\eta^{abc}\fd_{bci}$. The Poisson brackets of the basic variables are
\begin{eqnarray}
& & \left\{\au_{a}^{i}(x), \au_{b}^{j}(y)\right\}=0\nonumber\\
& & \left\{\au_{a}^{i}(x),
\ad_{b}^{j}(y)\right\}=\frac{1}{4\tilde{e}}\left(e_{a}^{i}e_{b}^{j}-2e_{a}^{j}
e_{b}^{i}\right)\delta^{3}(x,y)\label{21}\\
& & \left\{\ad_{a}^{i}(x), \ad_{b}^{j}(y)\right\}=0\nonumber
\end{eqnarray}
where $e_{a}^{i}\equiv \ad_{a}^{i}-\au_{a}^{i}$ and $\tilde{e}\equiv
\det e_{a}^{i}$.
Finally the constraints are given by
\begin{eqnarray}
& & \epsilon_{ijk}e_{a}^{j}\bu^{a}_{k}=0\label{22}\\
& & \epsilon_{ijk}e_{a}^{j}\bd^{a}_{k}=0\label{23}\\
& & e_{a}^{i}\bu^{a}_{i}=0\label{24}
\end{eqnarray}
If we define now the $3\times 3$ matrices $\ju_{ij}$ and $\jd_{ij}$
\begin{eqnarray}
& & \ju_{ij}\equiv e_{ai}\bu^{a}_{j}\label{25}\\
& & \jd_{ij}\equiv e_{ai}\bd^{a}_{j}\label{26}
\end{eqnarray}
we find that all the constraints are solved if we take a symmetric
$\jd_{ij}$ and a symmetric and traceless $\ju_{ij}$! This means, that by
coordinatizing the phase space by using the 18 variables per space point
given by $\ju_{ij}(x)$ and $\jd_{ij}(x)$, instead of the two connections
$\au_{a}^{i}$ and $\ad_{a}^{i}$ it is possible to solve the constraints is
a trivial way. The problem is that in order to effectively use this
formulation we should write the evolution equations in terms of the fields
$\ju_{ij}(x)$ and $\jd_{ij}(x)$ (since we know the action of the
diffeomorphism and vector constraints it would suffice to be able to write
the time evolution of $\ju_{ij}(x)$ and $\jd_{ij}(x)$ in terms of
themselves). This seems to require the inversion of the equations
(\ref{25}) and (\ref{26}); i.e. writing the connections in terms of
$\ju_{ij}(x)$ and $\jd_{ij}(x)$, a task not yet completed and probably
difficult. This is not a problem in any of the formulations discussed
above; in all of them it is straightforward to write the evolution
equations is terms of the relevant fields $\Psi_{ij}$ and
$\tilde{E}_{i}^{a}$ (or $A_{ai}$).

Another issue in the two-connection
formulation as compared to the $SO(3)$-ADM approach discussed in this
letter is the implementation of the reality conditions. The formulation
given by (\ref{22}-\ref{24}) is valid only for Euclidean space-times or
complex gravity. We must either impose reality conditions (and this
requires also knowing $\au_{ai}$ and $\ad_{ai}$ in terms of $\ju_{ij}$ and
$\jd_{ij}$) or use a modified Hamiltonian constraint and real fields.
Although a suitable Hamiltonian constraint for Lorentzian signature
space-times is known in terms of two real connections \cite{fer2}, it is
not straightforward to write it as a function of $\ju_{ij}(x)$ and
$\jd_{ij}(x)$ and so the simplicity of the solution to the constraints
discussed above for the Euclidean case is lost. This two connection
formulation may be useful to get some information about the relationship
between the vector constraint and the Gauss law because the constraints
generating three dimensional diffeomorphisms and internal $SO(3)$
rotations (\ref{22}) and (\ref{23}) have the same structure (they are
symmetric under the interchange of $\au_{ai}$ and $\ad_{ai}$). This, in
turn, may help to explain why we can either solve the vector constraint or
the Gauss law by using the Capovilla-Dell-Jacobson method in the different
Hamiltonian formulations.

In conclusion, the solution to the constraints of the $SO(3)$-ADM
formalism presented in this letter seems to provide a convenient way to
solve some of the constraints of the Hamiltonian formulation of general
relativity. In our opinion it has several advantages with respect to
similar approaches in the several Hamiltonian formulations presented
above. There is no need to implement any reality conditions
neither for Euclidean nor Lorentzian signatures, the constraint that is left
to solve gives a polynomial equation involving $\Psi_{ij}$ and
$\tilde{E}^{a}_{i}$, and the evolution equations can be written in a
straightforward way in terms of this fields.

{\bf Acknowledgements} I wish to thank A. Ashtekar, G. Mena, P. Peld\'an
and L. Smolin for several discussions and the Spanish Research Council
(CSIC) for providing financial support.

\end{document}